\documentclass[iop,twocolumn]{emulateapj}

\usepackage{amsmath}
\usepackage{amssymb}
\def\alphastar{\alpha_{{\rm star},i}-\alpha_{{\rm QSO},i}}

\begin{document}
\shorttitle{\sc Sgr Stream kinematics}
\shortauthors{ Koposov, Belokurov \& Evans }
\title{Sagittarius stream 3-D kinematics from SDSS Stripe 82}
\author{Sergey E. Koposov\altaffilmark{1,2},
	Vasily Belokurov\altaffilmark{1},
	N. Wyn Evans\altaffilmark{1}} 

\altaffiltext{1}{Institute of Astronomy, Madingley Road, Cambridge CB3 0HA, UK}
\altaffiltext{2}{Moscow MV Lomonosov State University, Sternberg Astronomical
Institute, Moscow 119992, Russia} 

\begin{abstract}
Using multi-epoch observations of the Stripe 82 region done by Sloan
Digital Sky Survey, we measure precise statistical proper motions of
the stars in the Sagittarius stellar stream. The multi-band photometry
and SDSS radial velocities allow us to efficiently select Sgr members
and thus enhance the proper motion precision to $\sim$ 0.1\,mas\,yr$^{-1}$. We measure separately
the proper motion of a photometrically selected sample of the main
sequence turn-off stars, as well as of a spectroscopically selected
Sgr giants. The data allow us to determine the proper motion
separately for the two Sgr streams in the South found in
\citet{Ko12}. Together with the precise velocities from SDSS, our
proper motion provide exquisite constraints of the 3-D motions of the
stars in the Sgr streams.
\end{abstract}

\keywords {Galaxy: halo, stars: kinematics, methods: statistical, surveys}

\section{Introduction}

The disintegrating Sagittarius (Sgr) dwarf galaxy remains a riddle,
wrapped in a mystery, inside an enigma.

Large scale photometric surveys, such as the Two Micron All-Sky Survey
(2MASS) and the Sloan Digital Sky Survey (SDSS) have now revealed the
structure of the tidal tails of the Sgr over more than $2\pi$ radians
on the Sky~\citep{Ma03,Be06}. By tallying all the stellar debris in
the streams and remnant, we now know that the progenitor galaxy had a
luminosity of $\sim 10^8 L_\odot$, comparable to the present day Small
Magellanic Cloud ~\citep{Ni10,Ni12}. The ingestation of such a large
progenitor, together with its dismantling under the actions of the
Galactic tides, can provide us with a wealth of information about both
the Galaxy and the Sgr, if we can only decode it.

Radial velocities, and sometimes metallicities and chemical
abundances, are now known for many hundreds of stars in the Sgr
tails~\citep[e.g.,][]{Ma04,Mo07,Ch07,Ch10}. There are also $\sim$ 10 
globular clusters associated with the Sgr tails~\citep{La10b}. This
rich mosaic of positions and velocities of Sgr tracers has proved
surprisingly difficult to understand.  Although there is no shortage
of Sgr disruption models in the literature~\citep[see
e.g.,][]{He04,La05,Jo05,Fe06}, they all have significant shortcomings, and fail
to
reproduce a substantial portion of the datasets. The most successful
recent attempt is by \citet{La10a}, though they do not explain the
striking two stream morphology seen in the SDSS data~\citep{Be06,
  Ko12,Sl13}. Additionally they advocate the use of a triaxial halo for the
Galaxy with minor axis contained with the Galactic plane, which is
unattractive on other grounds~\citep[e.g.][]{Ku94}. Given the impasse,
it is natural to look to proper motions of the Sgr stream as so as to
obtain a clearer picture of its space motion.

\cite{Ca12} have provided the first measurements of the proper motion
of the Sgr trailing tail. They took advantage of archival photographic
plate data in some of Kapteyn's Selected Areas which provides a 90
year baseline. They derive proper motions for four $40^\prime \times
40^\prime$ fields covering locations on the trailing tail between
$70^\circ$ and $130^\circ$ from the Sgr core. However, the number of
stars in each field remains modest ($\sim 15-55$), and so the
precision of the proper motion measurement is still quite low ($\sim
0.2-0.7$ mas yr$^{-1}$).

Here, we will pursue a different tack to obtain proper motions of the
trailing stream in roughly the same area of sky. As part of a project
to detect supernovae, the Sloan Digital Sky Survey scanned a $\sim
290$ square degree region on the Celestial Equator, known as Stripe
82~\citep[e.g.,][]{Ab09}. 
{ Proper motions can be derived by matching
objects between the $\sim 80$ epochs~\citep{Br08} obtained over time period of
$\sim$ 7 years}, whilst the co-added
optical data is roughly 2 magnitudes deeper than a single epoch SDSS
measurement. Although the baseline is small so the precision of a
measurement of proper motion of a single star is still low, we can
take advantage of the large number of Sgr tracers to get a high
precision {($\sim 0.1$\,mas\,yr$^{-1}$)} measurement for the proper motion of the
ensemble.
 
The paper is arranged as follows. Section 2 describes the extraction
of proper motions for stars from the Stripe 82 data using background
quasars to provide an absolute reference frame. Section 3 shows how to
identify the Sgr stars in Stripe 82, where they occupy a distinctive
niche in magnitude and radial velocity space. Section 4 discusses our
modelling of the proper motions using both photometric and
spectroscopic samples. We extract the proper motion for both the
bright and faint Sgr streams identified by \citet{Ko12}.  In section
5, we compare our proper motions both with the earlier work of
\citet{Ca12} and with the simulation data.

\section{Stripe 82 Proper Motion Determination}

\label{sec:pms}

Stripe 82 has already been the subject of numerous studies. Its
multi-epoch and multi-band imaging allows study of the variable sky
and identification of many kinds of transient phenomena~\citep[see
  e.g.,][]{Se07,Be08,Ko09,Wa09}.  The Stripe 82 dataset has also been
used to derive proper motions by \citet{Br08}. This light-motion
catalogue was subsequently exploited to build reduced proper motion
diagrams~\citep{Vi07} and analyse kinematical properties of Galactic
disk and halo populations \citep{Sm09a,Sm09b,Sm12}.

Even so, the proper motion measurements pioneered by \citet{Br08} can
be improved. For the bulk proper motion of Sgr, we are interested in
the statistical properties of a large ensemble of faint tracers, so
proper motions with the smallest possible systematic errors are highly
desirable. The original catalogue by Bramich does not provide proper
motions for stars fainter than $r \sim$ 20.5 and is known to have some
noticeable systematics. Since the Sgr stream has a very large number
of tracers in Stripe 82~\citep{Wa09}, we do not require a proper motion
measurement for every star, but rather need small systematic errors
and well understood error-bars for an ensemble. For this purpose, it
makes sense to measure the proper motions relative to quasars (QSOs).

Stripe 82 has a number of both spectroscopically and photometrically
identified QSOs. In this work, we have used the catalogue of
spectroscopic QSOs from \citet{Sc10} and the sample of photometrically
identified QSOs from \citet{Ri09} to extract denizens of Stripe
82. The purity of the spectroscopic catalogue of QSOs is guaranteed 
--
all the objects are QSOs and must have zero proper motion. However,
the photometric catalogue is known to have some contamination by
stars. In order to minimize contaminants, we use the cut {\tt good}$
\ge 1$, as recommended by \citet{Ri09}. This guarantees a small
stellar contamination, certainly $<$5\%.

\subsection{Relative Proper Motions}

Given a sample of QSOs each with zero proper motion, then for each
star in the vicinity of the QSO, we may determine proper motion
relative to the quasar.  As an input catalog for the stars, we took
the Stripe 82 co-add dataset \citep{An11}, from which we select
primary objects, classified by the SDSS pipeline as stars. { The
  individual source detections are taken from the Stripe 82 portion of
  the SDSS DR7 database \citep{Mu05} using only those fields having
  {\tt acceptable} and {\tt good} data quality flags. Matching
  co-added sources to detections at individual epochs is done with the
  0.5 arcsec radius using the Q3C module for the PostgreSQL database
  \citep{Ko06}.} This procedure makes the catalogue incomplete for
high proper motion objects (with proper motions $\gtrsim$ 100 mas
yr$^{-1}$), but we are not interested in such objects in our current
study.

Then, for each pair (star, QSO) observed multiple times by SDSS within
one field, we analyse the positional offsets and errors. For the right
ascension, these are defined via
\begin{eqnarray}
\Delta_i &=& \alphastar, \nonumber\\
\sigma_{\Delta,i} &=& (\sigma_{\alpha,{\rm
    star},i}^2+\sigma_{\alpha,{\rm QSO},i}^2)^{1/2}
\end{eqnarray}
with similar equations for the declination. The model for the
positional offsets is
\begin{eqnarray}
& & P(\Delta|t,\Delta_0,\mu,\sigma,f) =   f\, R(\Delta) + \nonumber\\
& &
\frac{1-f}{\sqrt{2\pi}\sigma}\exp\left(-\left(\frac{\Delta-\Delta_0-\mu\,t}{\sigma}\right)^2\right)
\label{eq:offsets}
\end{eqnarray}
where $\mu$ is the proper motion of the star, $t$ is the date of the
observation, $f$ is the fraction of outliers, and $\sigma$ is the
scatter around the linear relation, whilst $R(\Delta)$ is the
rectangular function to account for the outliers.  

The resulting likelihood is then minimized with respect to the 4
parameters $\mu$, $\Delta_0$, $f$, $\sigma$ with the error-bars
determined from the Hessian at the minimum. Repeating this procedure
for the offsets in declination gives us the proper motions and their
errors, $\mu_\alpha$, $\sigma_{\mu,\alpha}$, $\mu_\delta$,
$\sigma_{\mu,\delta}$ for all the sources with a spectroscopic or a
photometric QSO nearby.

\begin{figure*}
 \includegraphics{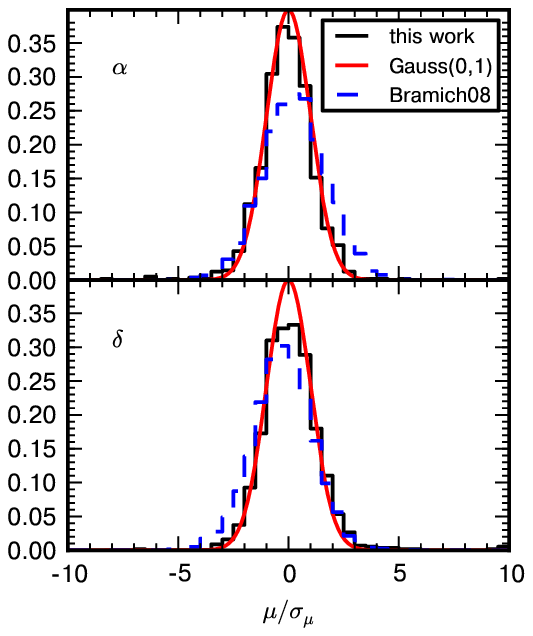}%}{plots/qso_errors}
 \includegraphics{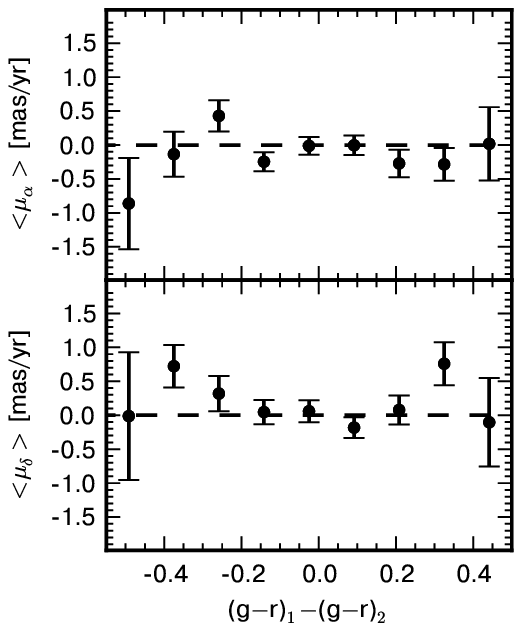}%{plots/qso_color_dep}
 \includegraphics{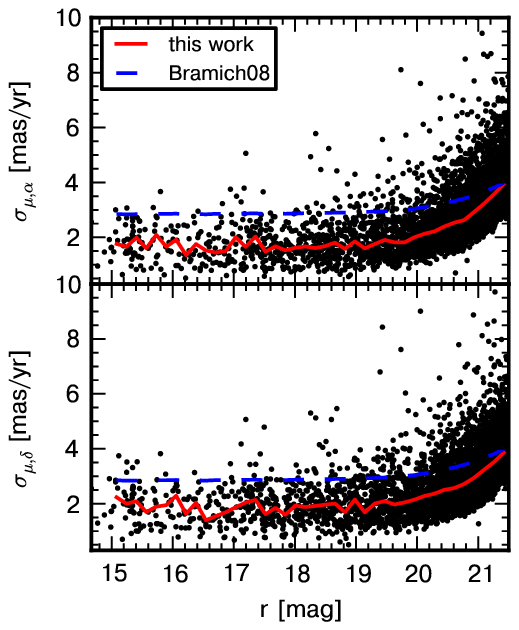}%plots/pm_prec}
\caption{ {\it Left panels:} Normalized histogram of proper motions of
  spectroscopic QSOs measured relative to the photometric QSOs
  normalized by the error bar provided by the modeling. Red lines are
  Gaussians with zero mean and unity dispersion. Blue dashed curves
  show the histograms of the proper motions of spectroscopic QSOs from
  \citet{Br08}. {\it Central panels:} The median proper motion of
  spectroscopic QSOs relative to the photometric QSOs versus their
  color-difference.  {\it Right panels:} The measured error bars on
  the the proper motions as a function of the $r$ band magnitude. The
  red lines show the median values of our measurement errors. The blue
  dashed lines show the median of proper motion errors from
  \citet{Br08}.}
\label{fig:qsocheck}
\end{figure*}

\subsection{Systematic Errors in the Proper Motions} 

After performing the computation of the proper motion for individual
stars, and before trying to measure the statistical proper motions for
ensembles of stars, it is important to check for the presence of
possible systematics, as well as to examine the accuracy of the error
bars.  Fig.~\ref{fig:qsocheck} presents such an assessment. In this
paper, we are focusing on the particular part of Stripe 82, which
intersects with the Sgr stream.  As systematic effects may depend on
the right ascension, Fig.~\ref{fig:qsocheck} uses only the proper
motions in the right ascension range $20^\circ<\alpha<50^\circ$.

The left panel of Fig.~\ref{fig:qsocheck} shows the histogram of
proper motions of the spectroscopic QSOs measured relative to the
photometric QSOs and normalized by the error bar provided by our
fitting procedure. The red overplotted curve shows a Gaussian with the
center at zero and unity dispersion. The excellent match of the
histogram with the Gaussian curve shows that that the proper motions
do not possess noticeable systematic offsets, and that the error bars
on the proper motions are a faithful description of the precision.  It
is also quite clear from the plot that this is not true for proper
motion measurements by \citet{Br08}, which have noticeable systematics
and error underestimation. The middle panel of Fig.~\ref{fig:qsocheck}
shows the the proper motion of the spectroscopic QSOs relative to the
photometric QSOs versus the $g\!-\!r$ color difference. The points
with error bars show the median proper motion in bins of $g\!-\!r$,
while the error bars are 1.48 times the median absolute deviation of
proper motions within a corresponding $g\!-\!r$ bin. Since all the
points lie within 1 $\sigma$ of zero, we conclude that the
color-dependent terms in the proper motions are negligible down to the
precision 0.1$-$0.2 mas yr$^{-1}$. This holds true at least within the
color range $-0.2\lesssim g\!-\!r\lesssim1$, which is the color-range
applicable to 99\% percent of the photometric QSOs. Later we will see
that the proper motion of the Sgr stream determined from the
photometric sample with $g\!-\!r\sim 0.3$, $g\!-\!i\sim 0.3$ and the
spectroscopic sample with $g\!-\!r\sim 0.5$, $g\!-\!i\sim 0.75$ agree each
other, further confirming the small level of color-related systematic
errors. And last, the right panel of Fig.~\ref{fig:qsocheck} shows the
precision of our proper motion measurements as a function of $r$-band
magnitude. There is a considerable scatter, but the median proper
motion precision of $\sim 2$\,mas yr$^{-1}$ is significantly better than that of
\citet{Br08} for
large range of magnitudes.

\begin{figure*}
\includegraphics{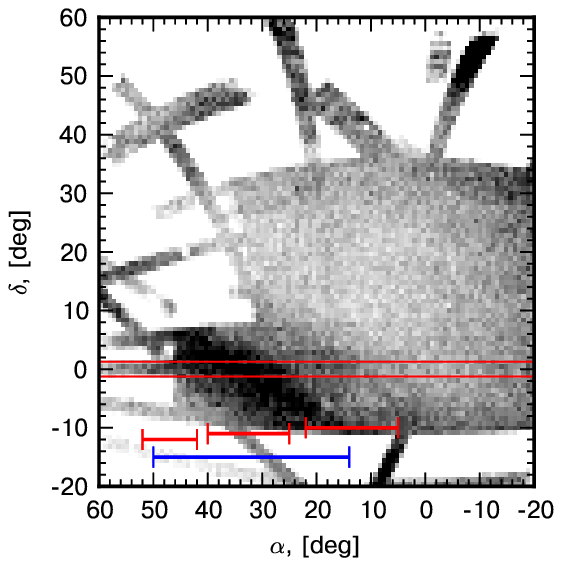}%{plots/arrows_plot0}
\includegraphics{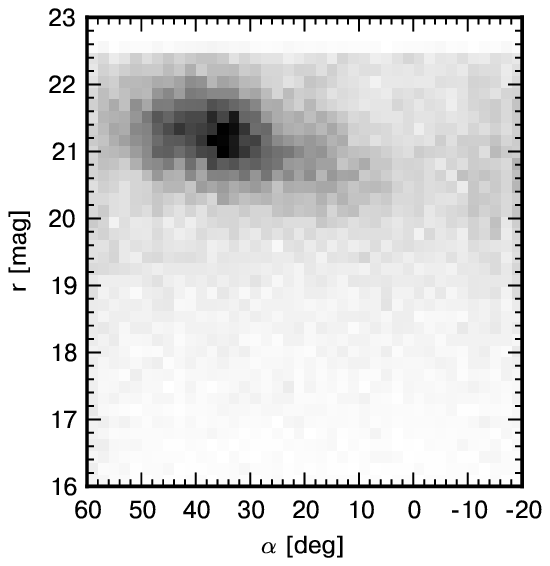}%{plots/ms_map}
\includegraphics{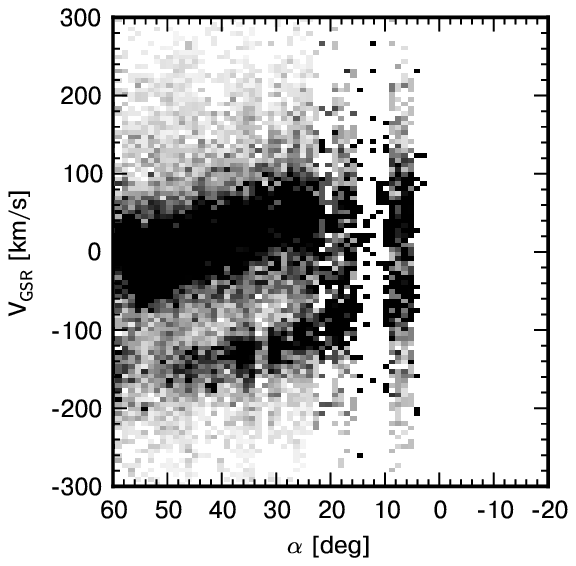}%{plots/ra_vel_sgr}
\caption{{\it Left panel:} Density of MSTO stars ($0.25<g\!-\!i<0.35$,
  $19.8 <r< 22$) from SDSS DR8 in the southern Galactic
  hemisphere. The SDSS Stripe 82 region is delineated by the red line.
  { Four horizontal error-bars show the range of right ascensions
    used to perform proper motion measurements in Stripe 82; red
    error-bars correspond to photometric samples, blue error-bars
    correspond to the spectroscopic sample.}  {\it Center panel:} The
  density of MSTO stars as a function of $r$ band magnitude and right
  ascension illustrating a distance gradient along the stream.  {\it
    Right panel} Radial velocities of stars near the Stripe 82. These
  are SDSS measurements of stars with $\log($g$)<4$ and located near
  the Sgr orbit $-15\degr<B<5\degr$ and near Stripe 82
  $|\delta|<20\degr$. The Sgr stream is obvious at radial velocities
  of $-200$\,km\,s$^{-1}<V_{\rm GSR}<-100\,$km\,s$^{-1}$. The change
  of radial velocity along the stream is also quite prominent.}
\label{fig:sag_map} 
\end{figure*}

\begin{deluxetable}{ccc}
\tabletypesize{\footnotesize}
%\tablewidth{0pt}
\tablecaption{Radial velocities and dispersions of Sgr stars along the stream
as traced by SDSS.}
\tablehead{ 
\colhead{$\Lambda$}  &
\colhead{V$_{\rm GSR}$} &
\colhead{$\sigma_{\rm V}$}   \\
\colhead{deg}  & 
\colhead{km\,s$^{-1}$} &
\colhead{km\,s$^{-1}$}
}
\startdata
87.5&  $-85.6 \pm  2.1$&  $13.7 \pm  2.1$\\
92.5&  $-100.5 \pm  1.4$& $ 14.8 \pm  1.5$\\
97.5&  $-106.0 \pm  1.5$&  $11.1 \pm  1.6$\\
102.5&  $-121.0 \pm 1.0$&  $13.0  \pm  1.0$\\
107.5&  $-132.5 \pm  1.2$&  $13.8 \pm  1.3$\\
112.5&  $-140.5 \pm  1.6$&  $19.5 \pm  1.4$\\
117.5&  $-146.7 \pm  1.5$&  $16.7 \pm  1.4$\\
125.0&  $-154.7 \pm  2.0$&  $19.9 \pm  2.8$\\
135.0&  $-156.8 \pm  2.8$&  $27.2 \pm  3.3$
\enddata
\label{tab:sgr_vel}
\end{deluxetable}

\section{The Sagittarius Stream}
\label{sec:sgr}

The Sgr stream in the southern Galactic hemisphere is known to have a
complex structure. \citet{Ko12} used main-sequence turn-off (MSTO)
stars extracted from SDSS Data Release 8 (DR8) to demonstrate the
existence of two streams -- a thicker, brighter stream and a thinner,
fainter stream displaced by $\sim 10^\circ$. The brighter stream has
multiple turn-offs as well as a prominent red clump, whereas the
fainter stream does not. \citet{Ko12} argued that the brighter stream
was composed of more than one stellar population, including some
metal-rich stars, whereas the fainter stream is dominated by a single
metal-poor population.
 
Here, we are primarily interested in the intersection of the Sgr
stream with the SDSS Stripe 82. Fig.~\ref{fig:sag_map} shows the
density of MSTO stars extracted via the cuts $0.25<g\!-\!i<0.35$ and
$19.8<r<22.5$ in the southern hemisphere, together with Stripe 82
demarcated by the red lines. Again, two distinct streams structures
are clearly visible -- a brighter one crossing the Equator and Stripe
82 at a right ascension of $\sim 35^\circ$ and a dimmer one crossing
the Equator at a right ascension of $\sim 15^\circ$. The magnitude
distribution of MSTO stars along Stripe 82 is shown in the middle
panel of Fig.~\ref{fig:sag_map}. The Sgr streams are clearly evident,
with the brighter stream visible at $\alpha\sim 35^\circ$ and fainter
one at $\alpha\sim 15^\circ$. Since the distances to the streams are
not constant with $\alpha$, and since Stripe 82 crosses the stream at
an angle, we observe a clear distance gradient with right ascension.

The right panel of Fig.~\ref{fig:sag_map} shows the radial velocity as
a function of right ascension for giant stars, extracted by the cut
$\log(g)<4$ using the SDSS spectroscopic measurements. This time not
only are stars within Stripe 82 included, but also those satisfying
$|\delta|<20^\circ$ and lying near the stream
($-15\degr<B<5\degr$). As already found by~\citet{Wa09}, the Sgr
stream is visible at the $V_{\rm GSR}\sim
-150$\,km\,s$^{-1}\ $\footnote{ Here and throughout the paper, we use
  $V_{LSR}$=235 km$^{-1}$ \citep{Bo09,Re09,De11} and a solar peculiar
  velocity of (U,V,W)=(-8.5,13.38,6.49)\,km\,s$^{-1}$ from
  \citep{Co11}}. The figure also shows the variation of radial
velocities as the stream stars return from the apocenter of the Sgr
orbit. As they will be needed later, we extract the radial velocities
along the stream, together with the velocity dispersion, by fitting a
Gaussian to the Sgr signal. { Although Figure~\ref{fig:sag_map}
  shows the radial velocities as a function of right ascension, it is
  important to realise that we perform these fits in bins along
  $\Lambda$, which is the angle along the stream from the remnant,
  measured positive in the trailing direction, as defined in
  \citet{Ma03}). The resulting measurements as a function of $\Lambda$
  are shown in Table~\ref{tab:sgr_vel}.}

Importantly, Fig.~\ref{fig:sag_map} demonstrates that the Sgr stream
stars are visible as a distinct bright feature in magnitude and
velocity space.  Therefore, this information can be used to select Sgr
member stars and statistically measure their properties such as proper
motion.

\begin{figure*}
\includegraphics[width=\textwidth]{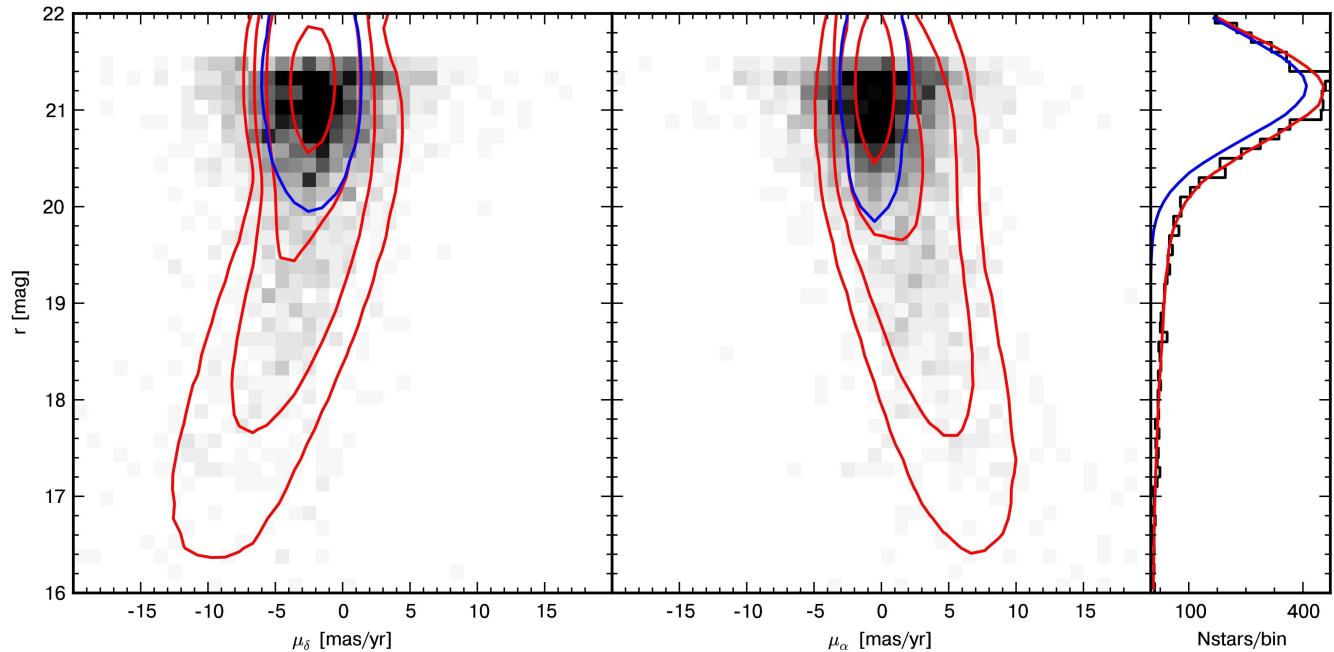}%{plots/r_pm}
\caption{ The bright Sgr stream component at
  $25^\circ<\alpha<40^\circ$ {\it Left and Middle panels:} Greyscale
  shows the 2D distribution of proper motions and magnitudes, while
  the red contours show the total error-deconvolved Gaussian mixture
  model. The component of the model corresponding to the stream is
  shown in blue. {\it Right panel:} 1D projection onto the apparent
  magnitude axis of the left panel.  The histogram of the data is
  shown in black, red curve shows the Gaussian mixture model of the
  luminosity function, whilst the model of the stream contribution is
  shown in blue.}
\label{fig:pm_hist}
\end{figure*}
\begin{figure*}
\includegraphics{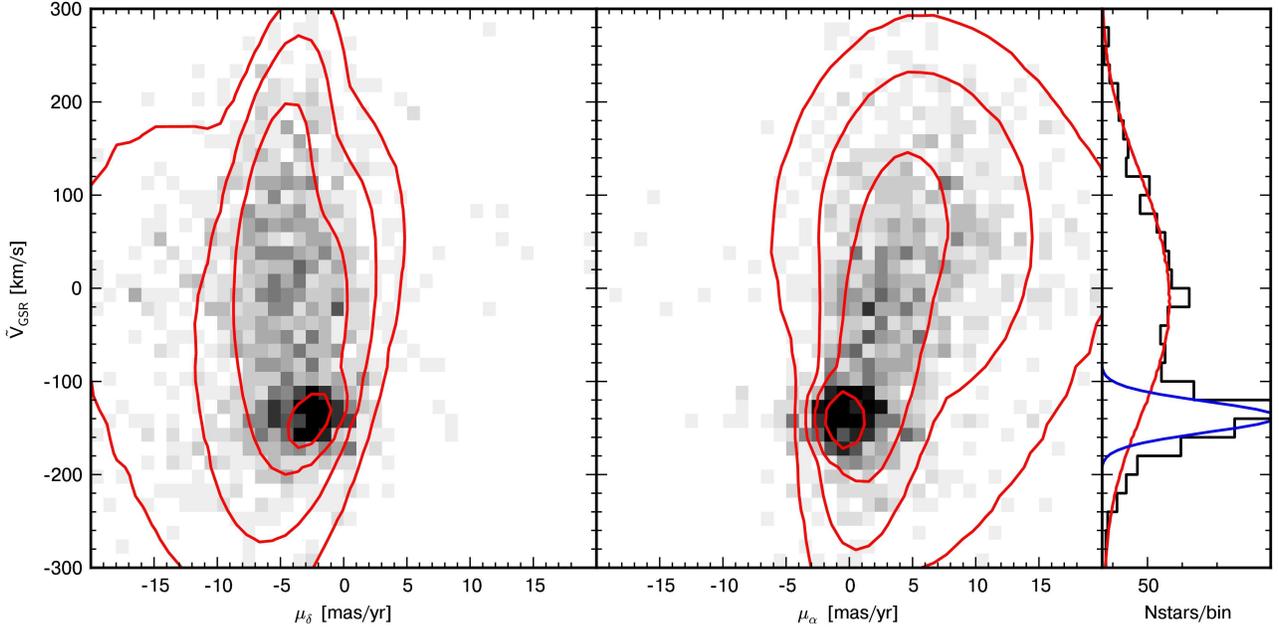}%{plots/vel_pm_fit2}
\caption{ {\it Left and Middle panels:} Greyscale shows 2D histograms
  of proper motions and the GSR radial velocities corrected for the
  stream's
  radial velocity gradient for all stars with SDSS spectra and
  $\log(g)<$4 in Stripe 82.  Red contours show the error-deconvolved
  Gaussian mixture model of the data.  {\it Right panel:} The histogram
  of the radial velocities. The Gaussian mixture model of the radial
  velocities is shown by the red curve and the stream component in
  blue.}
\label{fig:vel_pm_hist}
\end{figure*}
\begin{figure}
 \includegraphics{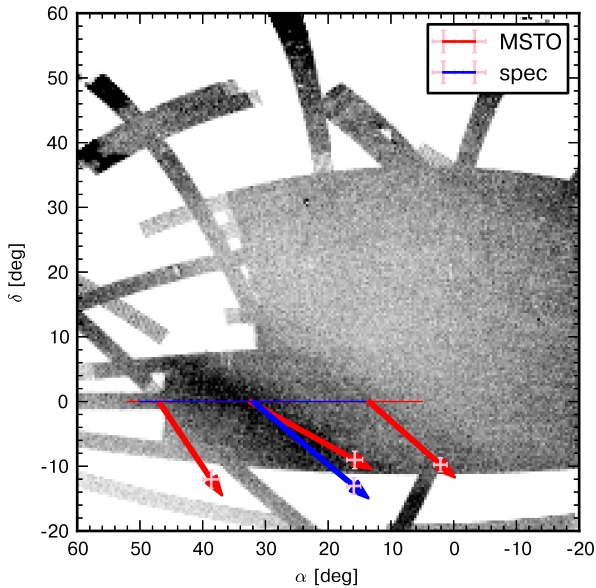}%{plots/arrows_plot}
\caption{Sgr streams in the South with the proper motion vectors
  overplotted.  The red vectors indicate the measurements performed
  using the photometrically selected MSTO stars at three different
  locations along the Stripe. Blue vector is for the spectroscopic
  sample. The measured error-bars of proper motions are shown in
  pink. The proper motion vectors have been corrected for the solar
  motion, assuming the distances from \citet{Ko12} and
$V_{LSR}=235$\,km\,s$^{-1}$ and peculiar velocity from \citet{Co11}.}
\label{fig:arrows_plot}
\end{figure}

\begin{deluxetable*}{ccccccccccc}
\tabletypesize{\footnotesize}
%\tablewidth{0pt}
\tablecaption{Proper motions measurements}
\tablehead{
\colhead{Field} 
&  \colhead{$\alpha_1$}
& \colhead{$\alpha_2$}
& \colhead{$\mu_\alpha\,cos(\delta)$\tablenotemark{b}}
& \colhead{$\mu_\delta$\tablenotemark{b}}
& \colhead{$\sigma_{\mu,\alpha\,cos(\delta)}$}
& \colhead{$\sigma_{\mu,\delta}$}
& \colhead{$\mu_l\,cos(b)$\tablenotemark{b}}
& \colhead{$\mu_b$\tablenotemark{b}}
& \colhead{$\sigma_{\mu,l}$}
& \colhead{$\sigma_{\mu,b}$} \\
\colhead{}
&  \colhead{deg}
& \colhead{deg} 
& \colhead{mas/yr}
& \colhead{mas/yr}
& \colhead{mas/yr}
& \colhead{mas/yr} 
& \colhead{mas/yr}
& \colhead{mas/yr}
& \colhead{mas/yr}
& \colhead{mas/yr}
}
\startdata
FP1 & 5.00 & 22.00 & 0.05 & -2.51 & 0.11 & 0.11 & 0.10 & -2.51 & 0.11 & 0.12\\
FP2 & 25.00 & 40.00 & -0.50 & -2.24 & 0.09 & 0.09 & 0.81 & -2.15 & 0.09 &
0.11\\
FP3 & 42.00 & 52.00 & 0.19 & -2.36 & 0.14 & 0.14 & 1.87 & -1.45 & 0.11 & 0.15\\
FS4\tablenotemark{a} & 14.00 & 50.00 & -0.42 & -2.65 & 0.11 & 0.13
&1.07 & -2.46 & 0.11 & 0.11
\enddata
\label{tab:pm_tab}
\tablenotetext{b}{not corrected for the Solar reflex motion}
\tablenotetext{a}{Spectroscopic sample} 
\tablecomments{ $\alpha_1$, $\alpha_2$ columns denote the edges of the boxes
in right ascension in Stripe 82 used to perform the proper motion measurements}
\end{deluxetable*}

\begin{deluxetable*}{ccccccccccccc}
\tabletypesize{\footnotesize}
\tablecaption{Positions and velocities of the Sgr stream.}
\tablehead{ 
  \colhead{Field}
&  \colhead{$\Lambda$} 
& \colhead{B}
& \colhead{X}
& \colhead{Y}
& \colhead{Z}
 & \colhead{Distance}
 & \colhead{U}
 & \colhead{$\sigma_U$}
 & \colhead{V}
& \colhead{$\sigma_V$}
& \colhead{W}
& \colhead{$\sigma_W$}
\\
\colhead{} 
& \colhead{deg}
& \colhead{deg} 
& \colhead{kpc}
& \colhead{kpc}
& \colhead{kpc}
& \colhead{kpc} 
& \colhead{km\,s$^{-1}$}
& \colhead{km\,s$^{-1}$}
& \colhead{km\,s$^{-1}$}
& \colhead{km\,s$^{-1}$}
& \colhead{km\,s$^{-1}$}
& \colhead{km\,s$^{-1}$}
}
\startdata
FP1 & 89.5 & 10.0 & -15.0 & 9.6 & -22.6 & 25.4 & 194 & 14 & -46 & 14 & 23 &
8\\
FP2 & 106.1 & 0.6 & -23.5 & 5.2 & -24.4 & 29.1 & 290 & 12 & 33 & 14 & -20 &
9\\
FP3 & 118.6 & -6.7 & -31.4 & 0.4 & -25.0 & 33.9 & 270 & 17 & -57 & 24 & -48
&
16\\
FS4\footnote{Spectroscopic Sample}   & 105.6& 0.8 & -23.2 & 5.3 & -24.3 &
28.9 & 308 & 15 & -16 & 18 & -40 & 11

\enddata

\label{tab:3dvel_tab}
\tablecomments{{ $\Lambda$, B, X, Y, Z correspond to the centers of the
fields where the proper motions are measured.}}

\end{deluxetable*}

\section{Modeling proper motions}

As demonstrated above, in the Southern Galactic hemisphere, it is
possible to achieve clean separation of the Sgr trailing tail stars
and the Galactic foreground in both apparent magnitude and the radial
velocity space.  { However, even for high-confidence stream
  members, the uncertainty of individual proper motion measurements
  ($\sim 2-3$\,mas yr$^{-1}$; see Section~\ref{sec:pms}) is comparable
  or higher than the expected tangential velocity of the stream
  \citep[$\sim 1-3$\,mas yr$^{-1}$;][]{La10a}.} It is therefore
crucial to combine the signal from as many stream members as possible
to beat down the noise. \citet{gd1} have shown that, for the regions
of apparent magnitude and radial velocity space dominated by the
stream, an accurate measurement of the systemic proper motion of an
ensemble of stars belonging to the stream can be obtained through
simple background subtraction.

Alternatively, the overall stellar distribution in the space of
observables can be modeled, yielding the direct contributions of the
Galaxy and the stream. If such models can be cast in the 3D space of
proper motion and magnitude or radial velocity, it is feasible that a
superior measurement of the proper motion can be achieved as the
contribution of the background to both systematic and random noise
will be reduced. Unfortunately, adequate analytical models of the Sgr
stream and the Galaxy are not readily available. Therefore, we choose
to approximate these distributions by a sum of multi-dimensional
Gaussians. This so-called Gaussian mixture is a well known
semi-parametric technique widely used to model multi-dimensional
datasets \citep{Mc00}. Gaussian mixtures have several key properties
that make the model-fitting straightforward and fast. First, the
uncertainties associated with the measurements (as well as the missing
data) are naturally incorporated into the model. Secondly, there
exists a guaranteed fast convergence procedure -- the Expectation
Minimisation (EM) algorithm \citep{De77}. In this work, we have used
the extreme-deconvolution package, the open-source Gaussian mixture
implementation by \citet{Bo11}.

\subsection{Photometric sample}
\label{sec:phot_sample}

As evident from the dissection of the SDSS dataset shown in
Fig.~\ref{fig:sag_map}, main sequence turn-off stars (MSTO) are the
most numerous Sgr stream specimens available to carry out the proper
motion analysis. The typical density of the MSTO stars in the stream
is 150 stars per square degree, compared to the foreground density of around 50.
Stripe 82 slices
through both the bright and the faint streams at an angle, resulting
in a slightly tilted bi-modal distribution in the plane of right
ascension and $r$ band magnitude. A faint Eastern wing to the bright
Stream can also be discerned at higher right ascension (see middle
panel of Fig.~\ref{fig:sag_map}).

Motivated by the evolution of the distance and the structure of the
stream along the Stripe, we split the MSTO sample into three parts:
the faint stream at $5^\circ<\alpha<22^\circ$, the bright stream at
$25^\circ<\alpha<40^\circ$, and the Eastern wing of the bright stream
at $42^\circ<\alpha<52^\circ$ { (we label those fields as FP1, FP2,
  FP3 respectively; they are shown by red error-bars on
Fig.~\ref{fig:sag_map})}. In each of the three right ascension bins, the
overall 3D distribution of MSTO stars in the space of
($\mu_\alpha,\mu_\delta,{\rm r}$) was modeled with a mixture of 5
Gaussians, The initial guess for the free parameters was obtained by
running the K-means algorithm \citep{Ma67}.  { The choice of number of Gaussians
  ($N_{\rm gau}$) was motivated by looking at the cross-validated
  log-likelihood \citep{Ar10} as a function of $N_{\rm gau}$. This
  initially rises as a function of $N_{\rm gau}$, peaks at $N_{\rm
    gau}\sim$ 5 and stays roughly constant for $N_{gau}>$5,
  demonstrating the goodness of fit and the absence of overfitting. We
  have also made sure that the objects in the Sgr stream are
  represented by a single Gaussian.}  When applying the extreme
deconvolution, we force the covariance matrix for the Sgr component to
be diagonal, e.g. assuming zero covariance between the proper motion
and the apparent magnitude. The proper motions for the stream stars
were not corrected for the solar reflex motion. The errors in the
proper motion measurement, i.e. the uncertainties in determining the
centre of the Gaussian representing the Sgr Stream, were determined
either from bootstrap procedure or from the Hessian matrix of the
likelihood function.

Fig.~\ref{fig:pm_hist} shows the data analysed as well as the best fit Gaussian
mixture model of the main Sgr stream $25^\circ<\alpha<40^\circ$. It is
reassuring to see that the Gaussian mixture model was able to describe
the data distribution adequately. The resulting measurements of the
proper motion together with the uncertainties are given in
Table~\ref{tab:pm_tab}. Among $\sim$ 7000 stars analyzed, according to the
model, $\sim$ 4000 stars belong to Sgr, while $\sim$ 3000 stars belong to the
background/foreground population.

\subsection{Spectroscopic sample}

{ The number of members of the Sgr stream with SDSS spectroscopic
  measurements is significantly smaller than the number of MSTO stars
  used in the previous section. Despite this, knowledge of the radial
  velocities and surface gravities allows us to have much purer
  samples of the Sgr stream members. In this section, we perform a
  complementary measurement of the stream's proper motion using
  spectroscopic members only.}
%Compared to the MSTO stars, the members of the stream with available
%SDSS spectra are significantly less numerous. 
%Nonetheless, using
%these, mostly giant, stars one can hope to leverage the purity of the
%sample to obtain a complementary measurement of the stream's proper
%motion of a comparable quality.

To this end, we select all stars with spectra in Stripe 82 lying at
$14^\circ<\alpha<50^\circ$ ({ we label this area FS4 and show it with the
blue error-bar on Fig.~\ref{fig:sag_map}}) and classified by the SDSS
spectroscopic
pipeline as giants $\log(g)<4$. Despite very wide range of
selected right
ascensions, most of the Sgr members with spectroscopy in the Stripe 82 region
located at the center of the bright stream, at $\alpha\sim30\degr$. According to
the right panel of Fig.~\ref{fig:sag_map}, the Sgr stream's radial velocity
changes along the Stripe. Therefore, to ease the modeling, we subtract the
variation of the radial velocity centroid from the data using the measurements
presented in Table~\ref{tab:sgr_vel}. The resulting radial velocity is
approximately constant as a function of RA, and the measurements of
proper motions and radial velocities $\mu_\alpha, \mu_\delta,
\tilde{V}_{\rm GSR} = V_{\rm GSR} - V_{\rm model,GSR}(\Lambda)$ can now be represented by a mixture of Gaussians.

We run the extreme deconvolution on the sample of $\sim$1500 stars and
find that 3 Gaussian components are sufficient to describe the
dataset. We follow the strategy outlined in Section
~\ref{sec:phot_sample}, with the difference that the covariance matrix
of the Gaussian representing the stream component is now set
free. Fig.~\ref{fig:vel_pm_hist} shows the density distribution of the
data, together with the best-fit Gaussian mixture model for the Stripe
82 stars with spectra. By comparing the grey-scale density with the
red contours, we can confirm that the model has captured the
properties of the dataset reasonably well. The last line of
Table~\ref{tab:pm_tab} reports the values of the proper motion for the
spectroscopic sample, and confirms that the measurements for the two
independent stream samples agree within 2$\sigma$. It is also
particularly reassuring since Sgr members in the spectroscopic sample
have very different colors (g$-$i $\sim$ 0.75) from the Sgr members in
the photometric sample (g$-$i $\sim$ 0.3) of the Sgr giants.
Therefore the agreement of the proper motions from two samples is a
proof of a small level of color-related systematic effects.

The proper motion signals we have measured have a very large
contribution of the solar reflex motion. Since we have distance
estimates to the Sgr stream in the South measured elsewhere
\citep[e.g.][]{Ko12}, we can correct for this and check whether the
proper motions are actually aligned with the
streams. Figure~\ref{fig:arrows_plot} shows the proper motion vectors
after applying the solar reflex corrections. As we can clearly see,
the proper motions are indeed properly aligned with the streams and are
consistent with each other. { For this calculation we have still 
used the distance to the faint stream as given by \citet{Ko12,Ko13},
although, there is evidence in \citet{Sl13} that the stream the faint
stream is 3$-$5 kpc closer.}

\section{Comparisons to earlier work}

There are many models of the Sgr stream in the literature, all
purporting to provide the distances, velocities and proper motions as
a function of position on the
sky~\citep[e.g.][]{Fe06,Pe10,La10a}. Although quite detailed, all the
models fail to reproduce at least some of the features that we see on
the sky~\citep[e.g.,][]{Ni10,Ko12}. But, it still instructive to see
where our data measurements lie relative to the existing models. We
have chosen the \citet{La10a} model as a comparison benchmark, because
it is arguably the most comprehensive and up-to-date.

There are several caveats to be borne in mind. The first is related to
the fact that while our central measurement in Stripe 82 at
$\alpha\sim35\degr$ or $(\Lambda,B)\sim(100\degr,0\degr)$ corresponds
directly to the center of the trailing tail in the simulation, the
fainter stream which crosses Stripe 82 at $\alpha\sim15\degr$ or
$\Lambda,B\sim(90\degr,10\degr)$ doesn't have a counterpart in the
simulation by \citet{La10a}. The second is related to the choice of
rotation velocity of the Local Standard of Rest. As shown in
\citet{Ca12}, the observed proper motion signal is sensitive to the
adopted V$_{\rm LSR}$. The models of \citet{La10a} have been computed
and fitted under the assumption of the IAU standard value of the
$V_{\rm LSR}$=220\,km\,s$^{-1}$, while the current best estimates are
slightly higher at $230-250$\,km\,s$^{-1}$\citep{Bo09,Re09,De11}. Throughout the
paper, we adopted the value of 235\,km\,s$^{-1}$. 

\begin{figure}
 \includegraphics{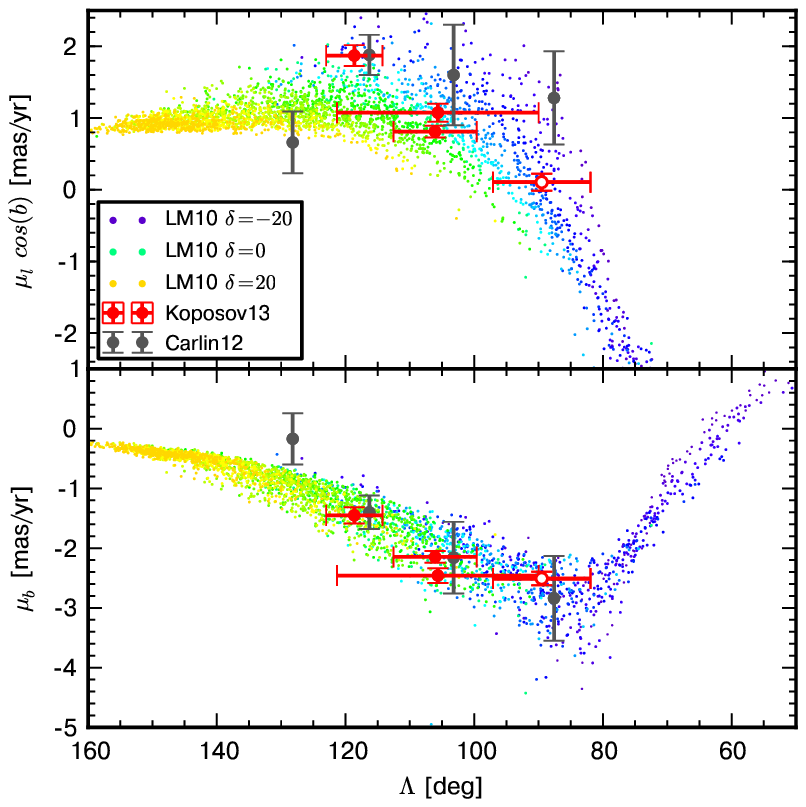}%{plots/law_maj_carl}
\caption{\label{fig:comparison}The comparison of the measured Stripe
  82 proper motions (red with error bars) with both the earlier
  measurements of \citet[][grey with error bars]{Ca12} and the
  simulations of \citet[][blue,green,yellow dots]{La10a}.  Our
  measurement of the proper motion of the fainter of the two trailing
  tails is identified by an empty red circle.  { Note that the
    datapoints from \citet{La10a} are selected from trailing tail
	 only
    and from the region near Stripe 82 $|\delta|<20\degr$. The points
    are colored according to $\delta$ such that points with $\delta
    \sim -20 \degr$ are dark blue, points with $\delta \sim 20\degr$
    are orange, and points near Stripe 82 are light-green}. Two of the
  fields from \citet{Ca12} lie within Stripe 82, while two other
  fields are $\sim$ 15 degrees away from the Stripe 82.}
\end{figure}

{ Fig.~\ref{fig:comparison} shows our data points overplotted onto
  the distribution of tracers from \citet{La10a} model, where we have
  selected tracers from the trailing tail only ($Lmflag=-1$ and $Pcol<=7$), and lying within 20
  degrees of Stripe 82 ($|\delta |<20$). The $\Lambda$ values of our
  measurements correspond to the centers of the rectangular areas used
  to perform the measurements, and the error-bars indicate the extent
  of those areas.}  The agreement for the central field ($\Lambda \sim
105\degr$), as well as for the edge of the bright stream ($\Lambda
\sim 120\degr$), is quite good. At the location of the fainter Sgr
stream (empty circle on the plot), the model doesn't have many
particles, as expected.  In order to facilitate further comparisons,
Table~\ref{tab:3dvel_tab} gives the positions and velocities
corresponding to our measurements. Here, ($X,Y,Z$) and ($U,V,W$) are
in a right handed Galactocentric coordinate system with the Sun
located at $X=-$8.5\,kpc, and $V_{\rm LSR}=235$\,km\,s$^{-1}$. The
distances taken from the work by \citet{Ko12}, and the error on the
velocities does not take into account any possible systematic error
distance of $\sim 10\%$. { Readers wishing to compare Sgr
  disruption models with our observations should use
  Table~\ref{tab:3dvel_tab} only for rough checks. A proper comparison
  entails measuring the average proper motions within the same area of
  the sky as we do and comparing these numbers directly with our
  Table~\ref{tab:pm_tab}.}

We also compare our measurements with the data from \citet{Ca12}.  Out
of 6 fields analysed by \citet{Ca12}, 4 have measurable Sgr signal and only 2
of those (SA 93 and SA 94)
lie near the Stripe 82. The other 2 fields with detectable Sgr signal are $\sim$
15 degrees away
from the Stripe 82. As Figure~\ref{fig:comparison} shows, the
error-bars of our measurements are much tighter than \citet{Ca12}, and
the proper motions themselves agree approximately within the combined
errors.

\section{Conclusions}

We have measured the proper motion of the Sgr stream using Stripe 82
data. By tying the astrometry to the known QSOs and combining the
measurements for large samples of stars, we have been able to achieve
high precision proper motions.  Our measurements have been performed
for two distinct groups -- spectroscopically selected red
giants/subgiants with SDSS radial velocities and photometrically
selected MSTO stars. The proper motions of those agree very well and
after correcting for the solar reflex motion are tightly aligned with
the Sgr stream. Our results are in agreement with earlier measurements
by \citet{Ca12}, but, by virtue of the large numbers of stars in our
samples, our statistical error bars are substantially smaller,
typically about $0.1$ mas yr$^{-1}$.

There are three fields in Stripe 82 for which the proper motion of the
photometric sample has been measured, and one field for the
spectroscopic sample. Combining this information with distances from
\citet{Ko12} gives us the full six dimensional phase space coordinates
of the Sgr trailing stream at four locations along Stripe 82.  We
provide a table of three dimensional positions and velocities of Sgr
stream stars, in which the contributions of the motion of the Sun and
LSR have been removed.

To complement the work on the Sgr streams in the South, it would be
particularly useful to carry out a corresponding kinematical analysis
for the streams in the North. This is a subject to which we plan to
return in a later contribution.  The combination of the proper
motions, radial velocities and distances in both Galactic hemispheres
should allow us to make further progress in the understanding of the
complicated structure of the Sgr streams and solve some of the riddles
posed by their existence.

\medskip
\acknowledgements{The authors would like to thank Jo Bovy for making
  his extreme deconvolution code available and supporting it. The
  extreme deconvolution code version used in this paper was r112. Most
  of the results presented in this paper have been done using open
  source software numpy/scipy/matplotlib and scikits-learn
  \citep{scikit-learn}. An anonymous referee helped us remove a number
  of obscurities from the paper.

    Funding for the SDSS and SDSS-II has been provided by the Alfred P.
Sloan Foundation, the Participating Institutions, the National Science
Foundation, the U.S. Department of Energy, the National Aeronautics and
Space Administration, the Japanese Monbukagakusho, the Max Planck Society,
and the Higher Education Funding Council for England. The SDSS Web Site is
http://www.sdss.org/.

The SDSS is managed by the Astrophysical Research Consortium for the
Participating Institutions. The Participating Institutions are the American
Museum of Natural History, Astrophysical Institute Potsdam, University of
Basel, University of Cambridge, Case Western Reserve University, University
of Chicago, Drexel University, Fermilab, the Institute for Advanced Study,
the Japan Participation Group, Johns Hopkins University, the Joint Institute
for Nuclear Astrophysics, the Kavli Institute for Particle Astrophysics and
Cosmology, the Korean Scientist Group, the Chinese Academy of Sciences
(LAMOST), Los Alamos National Laboratory, the Max-Planck-Institute for
Astronomy (MPIA), the Max-Planck-Institute for Astrophysics (MPA), New
Mexico State University, Ohio State University, University of Pittsburgh,
University of Portsmouth, Princeton University, the United States Naval
Observatory, and the University of Washington.

}

\end{document}